\documentclass[prd,twocolumn,amsmath,amssymb]{revtex4-1}
\usepackage{graphicx}
\usepackage{amsmath}
\usepackage{MnSymbol}
\usepackage{mathrsfs}
\usepackage{colortbl}
\makeatletter

\begin{document}

\title{Using a black hole to weigh light: \\ can the Event Horizon Telescope yield new information about the photon rest mass?}

\author{Robert P. Cameron}
\email{robert.p.cameron@strath.ac.uk www.ytilarihc.com}
\address{SUPA and Department of Physics, University of Strathclyde, Glasgow G4 0NG, U.K.}

\begin{abstract}
\noindent We point out that data collected by the Event Horizon Telescope or a similar project might yield new information about the photon rest mass $m_\gamma$, in the form of evidence that $m_\gamma\ne 0$ together with a lower bound on $m_\gamma$ or a new upper bound on $m_\gamma$. Using Sgr A*, there is scope to improve on the best upper bound obtained via laboratory tests of Coulomb's law by up to nineteen orders of magnitude.
\end{abstract}

\date{\today}
\maketitle


It is an exciting time for the study of black holes. The Event Horizon Telescope is due to release the first image of an event horizon, namely that of the black hole Sgr A* at the centre of our galaxy \cite{Balick74a, Brown82a}. This comes hot on the heels of the first direct detection of a binary black hole merger, using gravitational waves \cite{Abbott16a}. 

In this paper we point out that data collected by the Event Horizon Telescope or a similar project \cite{Issaoun19a} might yield new information about the photon rest mass $m_\gamma$.

Surprisingly, it is not yet known with certainty how much light weighs: it is usually assumed that $m_\gamma$ is \textit{exactly} zero, however this has never actually been confirmed. The most precise laboratory tests of Coulomb's law have only succeded in establishing that $0\le m_\gamma\lesssim10^{-50}\textrm{kg}$ \cite{Williams71a} and the strongest empirical claim made to date, on the basis of astronomical observations, is that $0\le m_\gamma\lesssim10^{-63}\textrm{kg}$ \cite{Chibisov76a} (the reliability of the assumptions underlying this claim has been disputed \cite{Adelberger07a}). The possibility remains that $m_\gamma \ne 0$.

The question of whether or not $m_\gamma$ is indeed exactly zero is of fundamental importance. If it were discovered that $m_\gamma$ has a non-zero value (no matter how small), Maxwell's equations would have to be replaced in principle by the Proca equations \cite{Proca36a, Goldhaber71a, Jackson01a}. This would dramatically alter our basic understanding of light: according to the Proca equations, the potential $A^\mu$ is uniquely defined and thus directly observable; light has three possible polarisation states, not two; light does not propagate at the universal speed limit $c$ but instead with phase and group speeds that differ from each other and vary with frequency, even in vacuum \cite{Goldhaber71a, Jackson01a}. A non-zero value for $m_\gamma$ would also imply the existence of a new elementary scalar field or fields to produce the mass \cite{Stueckelberg38a}.

A remarkable prediction was made by Bekenstein in 1971: a static black hole with electric charge has a Coulombic electric field outside the event horizon if $m_\gamma=0$ but \textit{no} electric field outside the event horizon if $m_\gamma\ne0$, regardless of how small $m_\gamma$ is; a corollary of the no-hair theorem \cite{Bekenstein72a, Dolgov07a}. Further static calculations revealed that the electric charge of matter outside the event horizon is, in effect, \textit{screened} by the hole if $m_\gamma\ne0$, with no such screening if $m_\gamma =0$ \cite{Vilenkin78a, Leaute85a, Dolgov07a}. Finally, dynamic calculations revealed that this screening is realised after a time $\tau \sim 1/\mu_\gamma c=\hbar/m_\gamma c^2$ \cite{Paul04a, Dolgov07a}, where $\mu_\gamma=m_\gamma c/\hbar$ is the Compton wavenumber of the photon. Thus, electromagnetic radiation produced by matter just outside the event horizon of a black hole should differ dramatically depending on whether $m_\gamma =0$ or $m_\gamma \ne 0$, assuming that this matter has been in the vicinity of the hole long enough for the screening to be realised if $m_\gamma \ne 0$. The possibility of exploiting this phenomenon to extract information about $m_\gamma$ does not appear to have been pointed out explicitly before, perhaps because of the extreme difficulty of the necessary observations.

We propose that data collected by the Event Horizon Telescope or a similar project \cite{Issaoun19a} be analysed with the above in mind. There are two distinct possibilities. If the electromagnetic radiation produced by matter just outside the event horizon of a black hole reveals that screening \textit{does} occur, it can be concluded that $m_\gamma \ne 0$ and, furthermore, that $m_\gamma>\hbar/\tau c^2$, where $\tau$ is the duration for which the matter has been in the vicinity of the hole. As indicated above, the discovery that $m_\gamma\ne0$ would be revolutionary. If, instead, the electromagnetic radiation produced by matter just outside the event horizon does \textit{not} reveal any sign of screening, it can be concluded that $0\le m_\gamma \lesssim\hbar/\tau c^2$. For Sgr A*, $\tau$ could conceivably be as large as the age of the Milky Way, in which case $\hbar/\tau c^2=10^{-69}\textrm{kg}$. Thus, there is scope to improve on the best upper bound on $m_\gamma$ obtained via laboratory tests of Coulomb's law ($10^{-50}\textrm{kg}$ \cite{Williams71a}) by up to nineteen orders of magnitude and even the strictest empirical upper bound on $m_\gamma$ yet claimed ($10^{-63}\textrm{kg}$ \cite{Chibisov76a}; disputed in \cite{Adelberger07a}) might be beaten, by up to six orders of magnitude.

This work was supported by The Leverhulme Trust (RPG-2017-048).



\end{document}